\documentclass[a4paper,fleqn,usenatbib]{mnras}

\usepackage[T1]{fontenc}
\usepackage{ae,aecompl}

\usepackage{graphicx}   
\usepackage{amsmath}    
\usepackage{amssymb}    
\usepackage{array}
\usepackage{epsfig}
\usepackage{lscape}
\usepackage{footnote}
\usepackage{etoolbox}
\usepackage{leftidx}
\usepackage{color}
\usepackage{cleveref}
\usepackage{caption}
\usepackage{subfig}
\usepackage{pdflscape}
\usepackage{threeparttable}
\usepackage{url}
\usepackage{letltxmacro}

\title[A ghostly DLA revealed]{A ghostly damped Ly$\alpha$ system revealed by metal absorption lines
\thanks{Based on data obtained  with XSHOOTER on the ESO-VLT; Prgm. 084.A-0699(A).}}

\author[H. Fathivavsari et al.]{
 H.~Fathivavsari,$^{1,2}$\thanks{E-mail: h.fathie@gmail.com} P.~Petitjean$^{1}$, S.~Zou$^{1}$, P.~Noterdaeme$^{1}$, C.~Ledoux$^{3}$,
 \newauthor 
 T.~Kr{\"u}hler$^{4}$ and
 R.~Srianand$^{5}$\\
$^{1}$Institut d'Astrophysique de Paris, Universit\'e Paris 6-CNRS, UMR7095, 98bis Boulevard Arago, 75014 Paris, France\\
$^{2}$School of Astronomy, Institute for Research in Fundamental Sciences (IPM), P. O. Box 19395-5531, Tehran, Iran \\
$^{3}$European Southern Observatory, Alonso de C\'ordova 3107, Casilla 19001, Vitacura, Santiago 19, Chile\\
$^{4}$Max-Planck-Institut f\"ur extraterrestrische Physik,
Giessenbachstra{\ss}e, D-85748 Garching, Germany\\
$^{5}$Inter-University Centre for Astronomy and Astrophysics, Post
Bag 4, Ganeshkhind, 411 007, Pune, India\\
}

\date{Accepted 000. Received 000; in original form 000}

\pubyear{2016}

\begin{document}
\label{firstpage}
\pagerange{\pageref{firstpage}--\pageref{lastpage}}
\maketitle

\begin{abstract}
We report the discovery of the first 'ghostly' damped Ly$\alpha$
absorption system (DLA), which is identified by the presence of
absorption from strong low-ion species at $z_{\rm abs}=1.70465$
along the line of sight to the quasar SDSSJ113341.29$-$005740.0 with
$z_{\rm em}=1.70441$. No Ly$\alpha$ absorption trough is seen
associated with these absorptions because the DLA trough is filled
with the leaked emission from the broad emission line region of the
quasar. By modeling the quasar spectrum and analyzing the metal
lines, we derive log$N$(H\,{\sc
i})(cm$^{-2}$)$\sim$21.0\,$\pm$\,0.3. The DLA cloud is small
($\le$\,0.32~pc) thus not covering entirely the broad line region
and is located at $\ge$~39~pc from the central active galactic
nucleus (AGN). Although the DLA is slightly redshifted relative to
the quasar, its metallicity ([S/H]=$-$0.41$\pm$0.30) is intermediate
between what is expected from infalling and outflowing gas. It could
be possible that the DLA is part of some infalling material
accreting onto the quasar host galaxy through filaments, and that
its metallicity is raised by mixing with the enriched outflowing gas
emanating from the central AGN. Current DLA surveys miss these
'ghostly' DLAs, and it would be important to quantify the statistics
of this population by searching the SDSS database using metal
absorption templates.
\end{abstract}




\section{Introduction}

Quasars are powered by the infall of gas into the gravitational
potential well of super-massive black holes, residing at the center
of distant galaxies. The infall of gas occurs, preferentially,
through the so-called {\it cold flows} along the filaments of the
cosmic web \citep{vandevoort2012} that are predicted by simulations
\citep{martin2016}. These cold flows of low metallicity gas can also
feed and trigger star formation activities in the host galaxies,
hence, enriching their interstellar medium \citep{almeida2015}.
Strong AGN-driven outflows, on the other hand, are required to
regulate and quench star formation activities, and also prevent
overgrowth of the galaxy (and the central black hole) by ejecting
the available material back into the intergalactic medium (IGM)
and/or circum-galactic medium (CGM; \citealp{silk1998,
cano-diaz2012, carniani2016}).

Luminous quasars are capable of launching powerful and energetic
outflows of gas up to very large distances. The pristine infalling
gas can therefore merge with the enriched outflowing material
anywhere from the innermost region of the AGN to the outermost
region of the halo of the host galaxy. When the infalling and
outflowing gas collide, they get mixed, shocked and compressed to a
high density, probably for a short period, before getting evaporated
by the incident energetic outflows \citep{namekata2014}. If a line
of sight to the central AGN passes through this shocked material,
one would expect to detect a strong H\,{\sc i} absorption in the
spectrum with $N$(H\,{\sc i})~$>$~10$^{17}$~cm$^{-2}$ at the
redshift of the quasar
\citep{vandevoort2012,finley2013,fathivavsari2015}.

Since the gas is compressed to a very small size, it may not fully
cover the broad emission line region (BELR) of the quasar.
Therefore, for any given neutral hydrogen column density, there
could be a combination of the cloud projected angular size and its
relative velocity from the AGN for which the partial coverage is
such that the leaked emission from the Ly$\alpha$ BELR could almost
fully fill the strong (or damped) Ly$\alpha$ absorption trough. If
this happens, one would not detect the corresponding Ly$\alpha$
absorption in the quasar spectrum. So far, no such combination of
DLA size and relative velocity (from the AGN) has ever been found
along quasars' sightlines. For the first time, in this letter, we
report the discovery of such a ghostly DLA cloud along the line of
sight to the quasar SDSS~J113341.29$-$005740.0 (here after
J1133$-$0057). The presence of this DLA is revealed through the
detection of absorption lines from neutral (e.g. C~{\sc i} or
Na~{\sc i}) and low-ion species associated with it. This absorber is
part of the \citet{ledoux2015} sample. These authors systematically
looked for C\,{\sc i} absorption in the Sloan Digital Sky Survey
(SDSS) quasar spectra in a search for cold gas at high redshift. We
show that the cloud is compressed to a very small size, and is
located very close to the central AGN. We argue that proximity to
the quasar's central engine, and small physical dimension are the
common characteristics of ghostly DLAs.

Ghostly DLAs are important objects as (by definition) they always
show signature of negligible covering factor of at least the
Ly$\alpha$ BELR, and probably probe the innermost regions of the
AGNs. Current surveys of DLAs miss this class of objects as their
employed techniques to find DLAs rely at least partly on the fact
that the residual flux at the bottom of the DLA reaches zero
intensity \citep{prochaska2004,noterdaeme2009}, which is not the
case in this new class of DLAs. Therefore, we might actually be
witnessing the tip of an iceberg as new techniques based on metal
absorption line templates would probably find more of these ghostly
DLAs in the SDSS spectroscopic database.

\section{Observation}

The spectrum of the quasar SDSS J1133$-$0057 was taken on 2011
January 7, using the XSHOOTER spectrograph \citep{vernet2011}
mounted on the Very Large Telescope in the course of a program to
search for cold gas using neutral carbon (see \citealp{ledoux2015}
for more details). The spectrum covers the wavelength range from
3000~\textup{\AA} to 2.5~$\mu$m with the nominal resolving power of
R~=~4350, 7450, and 5300 in the UVB, VIS, and NIR arms,
respectively, corresponding to slit widths of 1.0, 0.9 and 0.9
arcsec. We used the ESO pipeline v2.5.2 to reduce the raw spectrum.
The 11$^{''}$-long slit of the spectrograph provided us with enough
spatial pixels free of quasar emission, to perform optimal sky
subtraction. Arc lamp spectra were used for wavelength calibration,
and the spectrum of a standard star observed during the same night
was used to flux calibrate the spectrum.

\section{A ghostly DLA}

\begin{figure}
\centering
\begin{tabular}{c}
\includegraphics[bb=125 362 491 685,clip=,width=0.95\hsize]{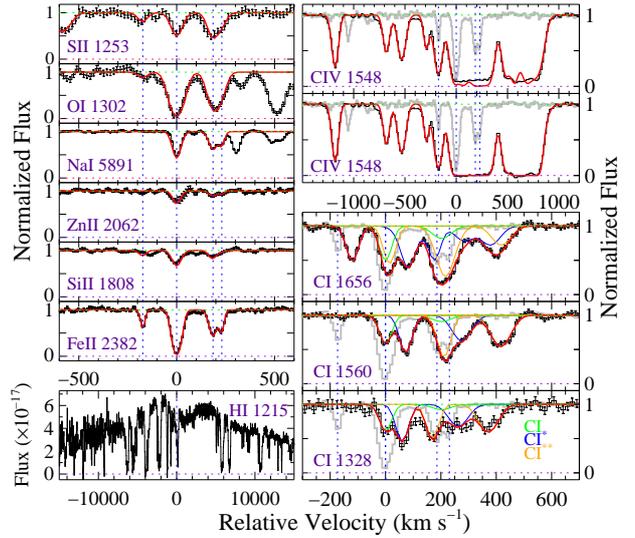}
\end{tabular}
\caption{Velocity plots and {\sc vpfit} solutions for the absorption
lines of the low and
high ion species detected in the ghostly DLA. The origin of the
velocity scale is at $z_{\rm abs}$~=~1.70465. The grey curve in the
C\,{\sc i} and C\,{\sc iv} panels is the Fe\,{\sc ii}\,$\lambda$2382
absorption profile. Note that no strong Ly$\alpha$ absorption is seen at the
redshift of the low ionization lines in the Ly$\alpha$ velocity
panel. The upper (resp. lower) C\,{\sc iv} panel shows the result
before (resp. after) taking the partial coverage effect into
account.}
 \label{j1133_1D}
\end{figure}

We detect a metal absorption line system in the spectrum of the
quasar J1133$-$0057, with an absorption redshift coincident with the
emission redshift of the quasar. The presence of species like
C\,{\sc i}, Cl\,{\sc i}, Mg\,{\sc i}, and especially Na\,{\sc i}
suggests that the absorber has a high neutral hydrogen column
density  and is probably a DLA \citep{petitjean2000}. However, as
shown in Fig.~\ref{j1133_1D}, no apparent DLA absorption line is
detected in the spectrum at the redshift of the low ion species,
i.e. $z_{\rm abs}$~=~ 1.70465. If one assumes that the Ly$\alpha$
BELR of the quasar is only partially covered by the DLA, then the
leaked Ly$\alpha$ emission from the BELR could fill the DLA
absorption trough and prevent direct detection. We show below that
this is indeed the case, and we demonstrate that the shallow
absorption dip seen around 3300~\textup{\AA} in the spectrum is
actually a ghostly signature of an otherwise strong DLA absorption
(see section 3.2). This shallow absorption feature allows us to
constrain the H\,{\sc i} column density of this elusive DLA.

\subsection{Quasar emission lines}

\begin{figure}
\centering
\begin{tabular}{c}
\includegraphics[bb=106 364 271 626,clip=,width=0.76\hsize]{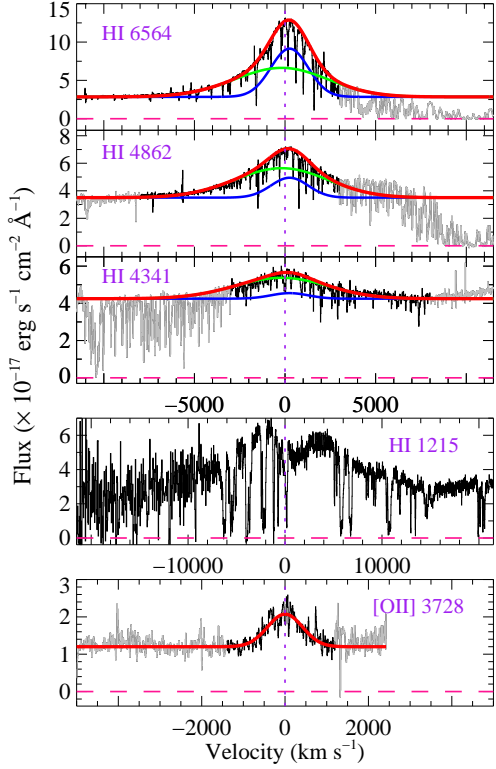}
\end{tabular}
\caption{Velocity plots and Gaussian function fits (where
applicable) to the H$\alpha$, H$\beta$, H$\gamma$, Ly$\alpha$ and
[O\,{\sc ii}] emission lines. The H-Balmer lines are fitted
simultaneously with two Gaussian components shown by the green and
blue curves in their corresponding panels. Note again that no DLA
absorption profile is seen in the H\,{\sc i}\,$\lambda$\,1215 panel.
The origin of the velocity plot is at $z_{\rm [O\,{\sc
II}]}$~=~1.70441. Regions shown in grey are excluded from the fit.}
 \label{hbalmeroii}
\end{figure}

We detect [O\,{\sc ii}]$\lambda$3728, Ly$\alpha$, H$\alpha$,
H$\beta$, and H$\gamma$  emission lines associated with the quasar,
as shown in Fig.~\ref{hbalmeroii}. A Gaussian function fitted to the
[O\,{\sc ii}] emission line doublet gives $z_{\rm
em}$~=~1.70441\,$\pm$\,0.00011, which is adopted as the systemic
redshift of the quasar. We simultaneously fit a 2-component Gaussian
function on the hydrogen Balmer emission lines, and obtain $z_{\rm
N}$~=~1.70668\,$\pm$\,0.00002, FWHM$_{\rm N}$~=~2344~km\,s$^{-1}$
and $z_{\rm B}$~=~1.70272\,$\pm$\,0.00005, FWHM$_{\rm
B}$~=~6048~km\,s$^{-1}$ as the redshift and FWHM of the narrow and
broad components of the fit, respectively (see
Fig.\ref{hbalmeroii}).

\begin{figure*}
\centering
\begin{tabular}{l}
\includegraphics[bb=43 436 521 613,clip=,width=0.80\hsize]{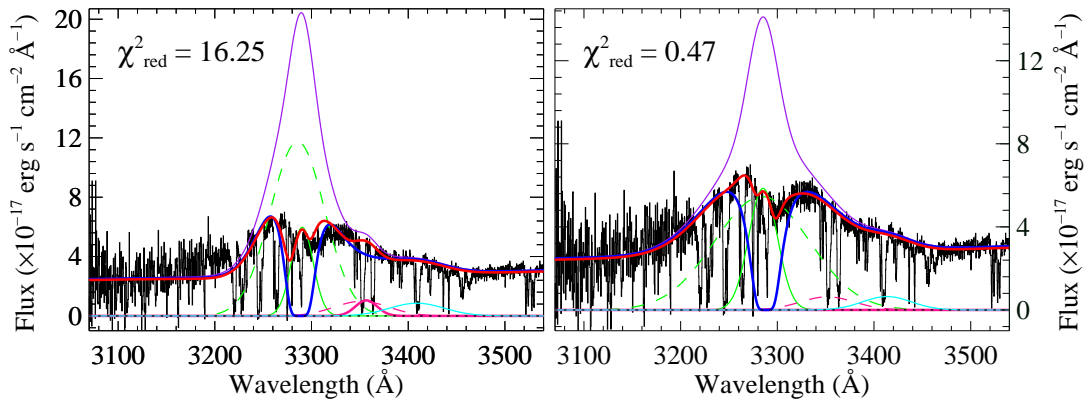}
\end{tabular}
\caption{Reconstruction of the observed spectrum around the
Ly$\alpha$ emission line spectral region. {\sl Left panel}: spectrum
corresponding to model(1) in Table~1. The dashed (resp.solid) green
and pink curves are the broad (resp. narrow) components of the BLR
Ly$\alpha$ and N\,{\sc v} emission lines, respectively. The cyan
curve shows the Si\,{\sc ii} emission line of the quasar, and the
purple curve is the combination of the quasar continuum and all
green, pink and cyan curves. The blue curve shows the combination of
the quasar continuum, the dashed green and pink curves, and a DLA
absorption profile with log\,$N$(H\,{\sc i})~=~21.1. Combining the
blue line with the solid green, pink and cyan curves gives the final
fit, which is overplotted on the observed spectrum as a red curve.
{\sl Right panel}: the lines are the same as in the left panel. In
contrast with the model shown in the left panel, in this model, the
redshifts and the widths of the broad and narrow components of the
Ly$\alpha$ BELR are free parameters (i.e. model(2) of Table~1). The
H\,{\sc i} column density in this model is log\,$N$(H\,{\sc
i})~=~21.0.}
 \label{dlamodels}
\end{figure*}

\subsection{Reconstructing the DLA absorption profile}

As mentioned earlier, the DLA absorption trough is almost completely
filled with the leaked emission from the BELR of the quasar. In this
section, we model the Ly$\alpha$ and N\,{\sc v} emission lines from
the BELR, and predict the amount of neutral hydrogen that is needed
to reproduce the shape of the observed spectrum in the Ly$\alpha$
spectral region. In our models, we assume that the Ly$\alpha$ and
N\,{\sc v} emission lines are each composed of a narrow and a broad
component, similar to what is seen for H-Balmer lines (see
section~3.1). During calculations, the redshift of the DLA is fixed
to the value obtained for the main component of the low-ion species
($z_{\rm abs}$~=~1.70465). Below, we briefly explain how the
reconstruction is performed.

First, we define the quasar's reddened continuum as a power law
function with $\alpha$~=~+1.5. The power law index, $\alpha$, is
obtained by fitting the quasar continuum observed at 3700, 4300, and
4660~\textup{\AA}. The broad component of Ly$\alpha$ (dashed green
curve in Fig.~\ref{dlamodels}) and N\,{\sc v} (dashed pink curve)
BEL are then added to this continuum to produce the
spectrum seen by the DLA cloud. The resulting spectrum after passing
through the DLA is shown as a blue curve in Fig.~\ref{dlamodels}.
Here, we assume that the covering factor of the broad component of
these emission lines is 1.0. The final simulated spectrum, which is
shown as a red curve in Fig.~\ref{dlamodels}, is obtained by adding
to the blue curve the narrow components of the Ly$\alpha$ and
N\,{\sc v} BEL profiles, which are assumed to have covering factor
of 0.0.

We first tried to reconstruct the observed spectrum with the maximum
number of fixed parameters. The result is shown in the left panel of
Fig.~\ref{dlamodels}, and the parameters of the fit are listed in
the third column of Table~\ref{dlamodelstable}. In this model, the
redshifts and the widths of the emission lines were fixed to those
of the H-Balmer lines. The H\,{\sc i} column density and the
amplitude of the emission lines are the only free parameters in this
model. The final fit, which is shown as a red curve in the left
panel of Fig.~\ref{dlamodels}, has a reduced chi-square of
$\chi^{2}_{\rm red}$~=~16.25 and does not match the observed
spectrum very well especially near the blue wing of the Ly$\alpha$
emission line. We, therefore, tried to construct a second model in
which we also relax the redshift and the width of the BLR Ly$\alpha$
emission line components. The result of this new model, which now
fits the observed spectrum more satisfactorily with $\chi^{2}_{\rm
red}$~=~0.47, is shown in the right panel of Fig.~\ref{dlamodels},
and the parameters of the fit are listed in the fourth column of
Table~\ref{dlamodelstable}. This model converges with
log\,$N$(H\,{\sc i})~=~21.0. We also checked that H\,{\sc i} column
densities smaller than 20.70 and larger than 21.30 results in fits
with large $\chi^{2}_{\rm red}$. Therefore, in this work, we adopt
log\,$N$(H\,{\sc i})~=~21.0\,$\pm$\,0.3. We emphasize that the
H\,{\sc i} column density is not very sensitive to the choice of the
free parameters, thanks to the presence of the shallow dip seen
around 3300\,{\textup{\AA}}, which strongly constrains the H\,{\sc
i} column density to be $\sim$\,10$^{21}$cm$^{-2}$.

 \begin{table}
 \centering
\caption{Parameters of the two models constructed to explain the
observed spectrum in the Ly$\alpha$ emission line spectral region.
Log\,$N$(H\,{\sc i}): H\,{\sc i} column density of the DLA. $z_{\rm
DLA}$: redshift of the strongest component of the low-ion species.
$F_{\rm Ly\alpha}$: peak flux of the broad component of the
Ly$\alpha$ emission line profile in
10$^{-17}$\,erg~s$^{-1}$\,cm$^{-2}$\,\textup{\AA}$^{-1}$. $Q_{\rm
Ly\alpha}$: ratio of the peak flux of the narrow component to that
of the broad component of the Ly$\alpha$ emission line profile.
$F_{\rm NV}$: peak flux of the broad component of the N\,{\sc v}
emission line profile, with the same unit as $F_{\rm Ly\alpha}$.
$Q_{\rm NV}$: ratio of the peak flux of the narrow component to that
of the broad component of the N\,{\sc v} emission line profile.
$z_{\rm Ly\alpha}^{\rm B}$ and $z_{\rm Ly\alpha}^{\rm N}$: redshifts
of the broad and narrow components of the Ly$\alpha$ emission.
$z_{\rm NV}^{\rm B}$ and $z_{\rm NV}^{\rm N}$: redshifts of the
broad and narrow components of the N\,{\sc v} emission. FWHM$_{\rm
Ly\alpha}^{\rm B}$ and FWHM$_{\rm Ly\alpha}^{\rm N}$: velocity
widths of the broad and narrow components of the Ly$\alpha$
emission. FWHM$_{\rm NV}^{\rm B}$ and FWHM$_{\rm NV}^{\rm N}$:
velocity widths of the broad and narrow components of the N\,{\sc v}
emission. All velocity widths are given in km\,s$^{-1}$. }
 \setlength{\tabcolsep}{14.0pt}
\renewcommand{\arraystretch}{1.05}
\begin{tabular}{c c c c}

\hline
  &   & Model(1) &  Model(2) \\
\hline


1    &   log\,$N$(H\,{\sc i})            &  21.1\,$\pm$0.30     &  21.0\,$\pm$0.30 \\

2    &   $z_{\rm DLA}$                   &  1.70465   &  1.70465  \\

3    &   $F_{\rm Ly\alpha}$      &  11.75     &  5.46 \\

4    &   $Q_{\rm Ly\alpha}$       &  0.51      &  1.07 \\

5    &   $F_{\rm NV}$            &  0.99      &  0.63 \\

6    &   $Q_{\rm NV}$             &  1.06      &  0.00 \\

7    &   $z_{\rm Ly\alpha}^{\rm B}$      &  1.70272   &  1.70159  \\

8    &   $z_{\rm Ly\alpha}^{\rm N}$      &  1.70668   &  1.70241  \\

9    &   $z_{\rm NV}^{\rm B}$            &  1.70272   &  1.70272 \\

10    &   $z_{\rm NV}^{\rm N}$           &  1.70668   &  1.70668 \\

11    &   FWHM$_{\rm Ly\alpha}^{\rm B}$     &  6048      &  10550   \\

12   &   FWHM$_{\rm Ly\alpha}^{\rm N}$      &  2344      &  3180     \\

13    &   FWHM$_{\rm NV}^{\rm B}$           &  6048      &   6048   \\

14    &   FWHM$_{\rm NV}^{\rm N}$           &  2344      &   2344   \\



\hline
\end{tabular}
 \label{dlamodelstable}
\end{table}

\subsection{Elemental abundances and depletion}

In this DLA, we detect metal absorption lines from O\,{\sc i},
O\,{\sc i}$^{**}$, Si\,{\sc ii}, Si\,{\sc ii}$^{*}$, C\,{\sc i},
C\,{\sc i}$^{*}$, C\,{\sc i}$^{**}$, C\,{\sc ii}, C\,{\sc ii}$^{*}$,
S\,{\sc ii}, Fe\,{\sc ii}, Zn\,{\sc ii}, Al\,{\sc ii}, Al\,{\sc
iii}, Mg\,{\sc i}, Mg\,{\sc ii}, Ca\,{\sc ii}, Na\,{\sc i}, Cl\,{\sc
i}, C\,{\sc iv}, Si\,{\sc iv}, and N\,{\sc v}. The velocity plots
and {\sc vpfit} \citep{carswell2014} solutions for some important
transitions are shown in Fig.~\ref{j1133_1D} and the derived total
column densities are listed in Table~\ref{coldens}. We used the same
technique as in \citet{fathivavsari2013} to fit the absorption
lines. We simultaneously fit five Fe\,{\sc ii} transitions to derive
the redshifts and the Doppler $b$-parameters of individual
components. The fits to the other low-ion transitions were conducted
by fixing the redshifts and the $b$-values to those obtained for
Fe\,{\sc ii}. The metallicity of the DLA is
[S/H]~=~$-$0.41\,$\pm$\,0.30 and [Zn/H]~=~$-$0.36\,$\pm$\,0.30. The
abundance of iron relative to sulphur and Zinc (i.e.
[Fe/S]~=~$-$1.93\,$\pm$\,0.10, [Fe/Zn]~=~$-$1.98\,$\pm$\,0.10)
suggests that the heavy elements are highly depleted onto dust
grains. Indeed, the quasar continuum is strongly reddened. Solar
photospheric abundances used here were taken from
\citet{asplund2009}.

\subsection{Partial coverage}
The flat-bottomed structure of the C\,{\sc iv} absorption profiles
suggests that they are saturated. However, as shown in
Fig.~\ref{j1133_1D}, the residual flux at the bottom of the profiles
does not reach zero intensity, implying that the C\,{\sc iv}
absorber may not be fully covering the C\,{\sc iv} BELR of the
quasar. In the upper C\,{\sc iv} panel in Fig.~\ref{j1133_1D}, we
show that it is not possible to fit the observed spectrum without
correcting for partial coverage. Indeed, after correcting for the
residual flux, we could successfully fit the observed spectrum (see
the lower C\,{\sc iv} panel in Fig.~\ref{j1133_1D}). The subtracted
residual flux amounts to about 7.5 per cent of the continuum and BLR
emission. This implies that $\sim$~19 per cent of the C\,{\sc iv}
BELR is not covered by the C\,{\sc iv} phase of the cloud.

In our spectrum, we also detect absorption from C\,{\sc i}
multiplets associated with the DLA in three different spectral
regions, with one multiplet (i.e. C\,{\sc i}\,$\lambda$1560) located
on the red wing of the quasar C\,{\sc iv} emission line. As shown in
Fig.~\ref{j1133_1D}, we could successfully conduct a 2-component fit
on the C\,{\sc i} multiplet transitions without invoking partial
coverage. This may imply that the C\,{\sc i}-bearing gas fully
covers the BELR of the quasar at the corresponding velocity.
However, we checked that even if we remove residual flux of up to 10
per cent from the bottom of the C\,{\sc i}\,$\lambda$1560 absorption
lines, we could still consistently fit the C\,{\sc i} multiplet
transitions, without increasing the $\chi^{2}_{\rm red}$ of the fit
by more than 5 per cent. Therefore, the observed C\,{\sc i}
multiplet absorption lines are {\it not} inconsistent with the
presence of partial coverage. Higher spectral resolution would allow
a deeper investigation of this issue.

 \begin{table}
   \begin{threeparttable}[b]
 \centering
\caption{Column densities (in cm$^{-2}$) of some important species
in logarithmic units. The uncertainty on the column densities from
the fits are about 0.1~dex. }
 \setlength{\tabcolsep}{3.1pt}
\renewcommand{\arraystretch}{1.5}
\begin{tabular}{c c c c c c c c c}

\hline

 S\,{\sc ii}   & Fe\,{\sc ii} & Zn\,{\sc ii} & Si\,{\sc ii} & O\,{\sc i} & Na\,{\sc i} & C\,{\sc i} & C\,{\sc i}$^{*}$ & C\,{\sc i}$^{**}$ \\

\hline 15.71 & 14.16 & 13.20 & 15.63 & $\ge$\,16.48
 & 12.94 & 14.30 & 14.76 & 14.74\\
 \hline

\end{tabular}
 \label{coldens}
                \end{threeparttable}
 \renewcommand{\footnoterule}{}
\end{table}

\section{Constraining the DLA location and size from fine structure lines}

Partial coverage of the Ly$\alpha$ BELR along this line of sight
implies that the cloud transverse size should be smaller than the
projected size of the BELR whose typical size is of the order of
$\sim$\,1~pc \citep{kaspi2005}. Below, we constrain the DLA size
using the physical state of the gas derived from the observation of
the C~{\sc i} fine structure transitions.

The observed column density ratios of the C\,{\sc i} multiplet
populations detected in the DLA are $N$(C\,{\sc
i}$^{*}$)/$N$(C\,{\sc i})~$\sim$~2.88 and $N$(C\,{\sc
i}$^{**}$)/$N$(C\,{\sc i})~$\sim$~2.75. The high values of these
ratios imply that the fine structure states are thermalized in the
cloud. As shown in figure~11 of \citet{jorgenson2010}, the level
populations at low densities (i.e. $n_{\rm HI}$~$<$~1000~cm$^{-3}$)
depend on the strength of the incident radiation field, $J_{\nu}$.
We extend their calculations to higher values of $J_{\nu}$ and find
that to match the observations, a radiation field with an intensity
of
$J_{\nu}$~=~1.4~$\times$~10$^{-16}$~erg~s$^{-1}$\,cm$^{-2}$\,Hz$^{-1}$\,sr$^{-1}$
is needed. On the other hand, if the density of the gas is high
enough (e.g. $n_{\rm HI}$~$>$~1000~cm$^{-3}$), then collisional
excitation alone will be sufficient to thermalize the system. In
this case, the presence of the incident radiation field will have no
further effect on the population of the fine structure states.

If $J_{\nu}$ were known, this would in principle allow us to
constrain the density, $n_{\rm HI}$. Here, we follow the technique
introduced by \citeauthor{wolfe2003} ({\color{blue}
2003} ; see also
\citealp{srianand2005}), using the C\,{\sc ii}$^{*}$ optical depth,
to estimate an upper limit to the intensity of the radiation field,
$J_{\nu}$, to which the DLA gas is exposed. Since C\,{\sc ii} and
C\,{\sc ii}$^{*}$ absorption lines are saturated in the spectrum, we
assume [C/H]~=~[S/H] and obtain log\,$N$(C\,{\sc ii})~=~17.02. We
further assume $N$(C\,{\sc ii}$^{*}$)/$N$(C\,{\sc ii})~$\le$~2.0
(see upper panel of figure~4 in \citealp{silva2002}) and get
log\,$N$(C\,{\sc ii}$^{*}$)~$\le$~17.32. The radiation field
intensity, $J_{\nu}$, is determined by assuming a steady state
condition and equating the heating rate, $\Gamma$, with the cooling
rate, $l_{c}$. The heating rate is given by
\begin{equation}
\Gamma = 10^{-5}\,\kappa\,\epsilon\,J_{\nu}~~~~~{\rm
erg~s^{-1}~H^{-1}}
\end{equation}
\noindent where $\kappa$ is the dust-to-gas ratio, $\epsilon$, the
grains heating efficiency, and $J_{\nu}$, the intensity of the
radiation field in erg~s$^{-1}$\,cm$^{-2}$\,Hz$^{-1}$\,sr$^{-1}$.
The dust-to-gas ratio is calculated using equation~7 of
\citet{wolfe2003}. With minimal depletion model, we get
$\kappa_{1}$~=~0.30 while maximal depletion model results in
$\kappa_{2}$~=~0.48. In this work, we adopt the mean value of
$\kappa$~=~0.39. Depending on the temperature of the cloud, the
grains heating efficiency, $\epsilon$, varies between 0.05 and 0.09
(\citealp{bakes1994, wolfire1995}, see their figure\,1). Adopting
$\epsilon$~=~0.07 as in \citet{wolfe2008} , and using equation~1, we
obtain the heating rate as
$\Gamma$~=~2.73~$\times$~10$^{-7}$\,$J_{\nu}$. The cooling rate, on
the other hand, is $l_{c}$~=~2.89$\times$10$^{-20}$$N$(C\,{\sc
ii}$^*$)/$N$(H\,{\sc i})~$\sim$~7.59\,$\times$\,10$^{-24}$
\,erg~s$^{-1}$~H$^{-1}$. By equating the cooling rate with the
heating rate, we get
$J_{\nu}\,\le$\,2.78\,$\times$\,10$^{-17}$\,erg~s$^{-1}$\,cm$^{-2}$\,Hz$^{-1}$\,sr$^{-1}$
as an upper limit to the intensity of the radiation field that is
impinging upon the DLA cloud. This value of $J_{\nu}$ is $\sim$\,5
times lower than what is needed to thermalize the C\,{\sc i} fine
structure levels (see above), implying that these fine structure
states are thermalized predominantly through collisional excitation.

With the above limit on $J_{\nu}$, the observed $N$(C\,{\sc
i}$^{*}$)/$N$(C\,{\sc i}) and $N$(C\,{\sc i}$^{**}$)/$N$(C\,{\sc i})
ratios can be reproduced {\it only} when $n_{\rm
HI}$~>~1000~cm$^{-3}$. This results in a characteristic size of the
cloud of $l$~=~$N$(H\,{\sc i})/$n_{\rm HI}$~$\lesssim$~0.32~pc. One
can also use the quasar luminosity, $L_{\nu}$
(\,=\,2.28\,$\times$\,10$^{27}$~erg~s$^{-1}$\,Hz$^{-1}$), to derive
from the upper value on $J_{\nu}$ a lower limit of $d$~$\ge$~66~pc
for the DLA-quasar separation.

Grains in the DLA clouds are mainly heated by photons with energy
$h\nu$~=~6$-$13.6~eV \citep{wolfe2003}. As these photons can reach
deeper into the cloud, they are absorbed by dust. When this
extinction by dust grains is taken into account, the lower limit to
the DLA-quasar separation decreases to $d$~$\ge$~39~pc.

\section{Summary and Conclusions}

We reported the discovery of the first ghostly DLA found along the
line of sight to the quasar SDSS J1133$-$0057. No Ly$\alpha$
absorption line is seen in the quasar spectrum at the redshift of
the detected low-ion species such as C\,{\sc i}, Na\,{\sc i},
Ca\,{\sc ii}, etc because the leaked emission from the Ly$\alpha$
BELR fills the DLA absorption trough. The DLA cloud is small
($\le$\,0.32~pc), and is located at $d$~$\ge$~39~pc from the central
AGN thus not covering entirely the broad line region.

Interestingly, this ghostly DLA has characteristics similar to those
of the 'eclipsing' DLA found along the line of sight to the quasar
J0823$+$0529 (Fathivavsari et al. 2015). Both DLAs have very small
sizes and are located close to the central AGN. However, in contrast
to the newly found ghostly DLA, the eclipsing DLA towards
J0823$+$0529 shows strong DLA absorption profile in the quasar
spectrum. This implies that the eclipsing DLA, with a transverse
size of 2.3\,$\le$\,$l$(H\,{\sc i})\,$\le$\,9.1\,pc, blocks most of
the radiation emitted by the quasar BELR. Because of the sub-parsec
size of the ghostly DLA, most emission from the quasar Ly$\alpha$
BELR is freely passing by the DLA and consequently filling the DLA
absorption trough, resulting in the observed spectral profile shown
in Fig.~\ref{dlamodels}. It could be possible that the two (ghostly
and eclipsing) DLAs are from the same population and are just two
stages of the same flow. Eclipsing DLAs could turn into ghostly DLAs
as strong outflowing material from the central AGN strikes and
compresses them into clouds of smaller size and higher density.
Statistical study of a large sample of ghostly and eclipsing DLAs
would in principle allow us to confirm whether there is any
connection between the two absorbers, and what importance they have
over the whole DLA population.














\bsp    
\label{lastpage}
\end{document}